\documentclass[12pt]{article}
\newcommand{\cent}{\centerline}
\newcommand{\h}{\hspace*{5 ex}}

\begin{document}
\vspace{0.5cm}

\cent{\bf ETTORE MAJORANA: HIS SCIENTIFIC (AND HUMAN) PERSONALITY}

\

\cent{{\bf E.Majorana: Scientist and Man}}

\

\

Erasmo Recami\\
        {\em Universit\`a statale di Bergamo, Bergamo, Italy;} and

        {\em INFN-Sezione di Milano, Milan, Italy.}

        e-mail: recami@mi.infn.it

\

{\bf abstract:} This article represents the English version of
part of the invited talk delivered by the author at the
``International Conference on Majorana's Legacy and the Physics of
the XXI century" held at the Department of Physics, University of
Catania, Catania, Italy, on 5 and 6 October, 2006: Conference
organized for celebrating the 100th anniversary of the birth (in
Catania) of the Italian theoretical physicist Ettore Majorana;
probably the brightest Italian theoretician of the XX century
(Enrico Fermi regarded him as the brightest in the world of his
time), even if to some people Majorana is still known chiefly for
his mysterious disappearance, in 1938, when he was 31. \ In this
writing we outline the significance of his main publications, as
well as his life: the biographical data being based on letters,
documents, testimonies discovered or collected by the author
during almost 30 years, and contained in the book by E.Recami,
``Il Caso Majorana: Epistolario, Testimonianze, Documenti"
(initially published, in 1987 and 1991, by Mondadori, Milan,
Italy, and presently published, in 2002, by Di Renzo Editore,
Rome, Italy: www.direnzo.it). \ At last, some information and
comments are added with regard to the scientific {\em manuscripts}
left unpublished by Majorana. \ [Obs.: Much of the material
following below is protected by copyright, and cannot be further
reproduced without the written permission of Fabio Majorana et
al., of E.Recami, and of the publisher Di Renzo].

\section{Historical Prelude.}

\h Ettore Majorana's fame solidly rests on testimonies like the following,
from the evocative pen of Giuseppe Cocconi. At the request of Edoardo
Amaldi\cite{Amaldi}, he wrote from CERN (July 18, 1965):

\

\h     ``In January 1938, after having just graduated, I was
invited, essentially by you, to come to the Institute of Physics at the
University in Rome for six months as a teaching assistant, and once I
was there I would have the good fortune of joining Fermi, Bernardini (who had
been given a chair at Camerino a few months earlier) and Ageno
(he, too, a new graduate), in the research of the products of disintegration
of $\mu$ ``mesons" (at that time called mesotrons or yukons), which are produced
by cosmic rays [...]

\h  ``It was actually while I was staying with Fermi in the small
laboratory on the second floor, absorbed in our work, with Fermi
working with a piece of Wilson's chamber (which would help to
reveal mesons at the end of their range) on a lathe and me
constructing a ``jalopy" for the illumination of the chamber,
using the flash produced by the explosion of an aluminum ribbon
short-circuited on a battery, that Ettore Majorana came in search
of Fermi.  I was introduced to him and we exchanged few words. A
dark face.  And that was it.  An easily forgettable experience if,
after a few weeks while I was still with Fermi in that same
workshop, news of Ettore Majorana's disappearance in Naples had
not arrived.  I remember that Fermi busied himself with
telephoning around until, after some days, he had the impression
that Ettore would never be found.

\h  ``It was then that Fermi, trying to make me understand the
significance of this loss, expressed himself in quite a peculiar way;
he who was so objectively harsh when judging people.  And so, at this
point, I would like to repeat his words, just as I can still hear them
ringing in my memory: `Because, you see, in the world there are various
categories of scientists: people of a secondary or tertiary standing,
who do their best but do not go very far.  There are also those of high
standing, who come to discoveries of great importance, fundamental for
the development of science' (and here I had the impression that he
placed himself in that category).  `But then there are geniuses like
Galileo and Newton.  Well, Ettore was one of them.  Majorana had what no
one else in the world had [...]'"

\

\h     And, with first-hand knowledge, Bruno Pontecorvo, adds: \ ``Some
time after his entry into Fermi's group, Majorana already possessed such an
erudition and had reached such a high level of comprehension of physics
that he was able to speak on the same level with Fermi about scientific
problems.  Fermi himself held him to be the greatest theoretical
physicist of our time.  He often was astounded [...].  I remember
exactly these words that Fermi spoke: `If a problem has already been
proposed, no one in the world can resolve it better than Majorana.' "
(See also \cite{Ponte}.)

\h  Ettore Majorana disappeared rather mysteriously on March 26,
1938, and was never seen again \cite{ER}. The myth of his
``disappearance" has contributed to nothing more than the
notoriety he was entitled to, for being a true genius and a genius
well ahead of his time.

\h Majorana was such a pioneer, that even his manuscripts known as
the {\em Volumetti}, which comprise his {\em study} notes written
in Rome between 1927, when he abandoned his studies in engineering
to take up physics, and 1931, are a paragon not only of order,
based on argument and even supplied with an index, but also of
conciseness, essentiality and originality: So much so that those
notebooks could be regarded as an excellent modern text of
theoretical physics, even after about eighty years, and a
``gold-mine" of seminal new theoretical, physical, and
mathematical ideas and hints, quite stimulating and useful for
modern research. Such scientific manuscripts, incidentally, have
been published for the first time (in 2003) by
Kluwer\cite{Kluwer}; and republished in 2006 in the original
Italian language\cite{Zanichelli}. But Majorana's most interesting
notebooks or papers --those that constituted his ``research notes"
will not see the light in the near future: it being too hard the
task of selecting, interpreting and...electronically typing them!
\ Each notebook was written during a period of about one year,
starting from the years ---as we said above--- during which Ettore
Majorana was completing his studies at the University of Rome.
Thus the contents of these notebooks range from typical topics
covered in academic courses to topics at the frontiers of
research.  Despite this unevenness in the level of sophistication,
the style in which any particular topic is treated is never
obvious. As an example, we refer here to Majorana's study of the
shift in the melting point of a substance when it is placed in a
magnetic field or, more interestingly, his examination of heat
propagation using the ``cricket simile." Also remarkable is his
treatment of contemporary physics topics in an original and lucid
manner, such as Fermi's explanation of the electromagnetic mass of
the electron, the Dirac equation with its applications, and the
Lorentz group, revealing in some cases the literature preferred by
him. As far as frontier research arguments are concerned, let us
here recall only two illuminating examples: the study of
quasi-stationary states, anticipating Fano's theory by about 20
years, and Fermi's theory of atoms, reporting analytic solutions
of the Thomas-Fermi equation with appropriate boundary conditions
in terms of simple quadratures, which to our knowledge were still
lacking.

\h   Let us recall that Majorana, after having switched to physics
at the beginning of 1928, graduated with Fermi on July 6, 1929,
and went on to collaborate with the famous group created by Enrico
Fermi and Franco Rasetti (at the start with O.M.Corbino's
important help); a theoretical subdivision of which was formed
mainly (in the order of their entrance into the Institute) by
Ettore Majorana, Gian Carlo Wick, Giulio Racah, Giovanni Gentile
Jr., Ugo Fano, Bruno Ferretti, and Piero Caldirola. The members of
the experimental subgroup were: Emilio Segr\'e, Edoardo Amaldi,
Bruno Pontecorvo, Eugenio Fubini, Mario Ageno, Giuseppe Cocconi,
along with the chemist Oscar D'Agostino. \ Afterwards, Majorana
qualified for university teaching of theoretical physics (``Libera
Docenza") on November 12, 1932; spent about six months in Leipzig
with W. Heisenberg during 1933; and then, for some unknown
reasons, stopped participating in the activities of Fermi's group.
He even ceased publishing the results of his research, except for
his paper ``Teoria simmetrica dell'elettrone e del positrone,"
which (ready since 1933) Majorana was persuaded by his colleagues
to remove from a drawer and publish just prior to the 1937 Italian
national competition for three full-professorships.

\h    With respect to the last point, let us recall that in 1937
there were numerous Italian competitors for these posts,
and many of them were of exceptional caliber; above all: Ettore Majorana,
Giulio Racah, Gian Carlo Wick, and Giovanni Gentile Jr. (the
son of the famous philosopher bearing the same name, and the inventor of
``parastatistics" in quantum mechanics). The judging committee
was chaired by E. Fermi and had as members E. Persico,
G. Polvani, A. Carrelli, and O. Lazzarino.  On the recommendation of the
judging committee, the Italian Minister of National
Education installed Majorana as professor of theoretical physics
at Naples University because of his ``great and well-deserved fame,"
independently of the competition itself; actually,
``the Commission hesitated to apply the normal university
competition procedures to him." The attached report
on the scientific activities of Ettore Majorana, sent to the minister by
the committee, stated:

\

\h     ``Without listing his
works, all of which are highly notable both for their originality of the
methods utilized as well as for the importance of the achieved results,
we limit ourselves to the following:

\h     ``In modern nuclear theories, the contribution made by this researcher
to the introduction of the forces called ``Majorana forces" is
universally recognized as the one, among the most fundamental,
that permits us to theoretically comprehend the reasons for nuclear
stability.  The work of Majorana today serves as a basis for the most
important research in this field.

\h     ``In atomic physics, the merit of having resolved some of the most
intricate questions on the structure of spectra through simple and
elegant considerations of symmetry is due to Majorana.

\h     ``Lastly, he devised a brilliant method that permits us to treat the
positive and negative electron in a symmetrical way, finally eliminating
the necessity to rely on the extremely artificial and unsatisfactory
hypothesis of an infinitely large electrical charge diffused in
space, a question that had been tackled in vain by many other
scholars."

\

One of the most important works of Ettore Majorana, the one that
introduces his ``infinite-components equation" was not mentioned,
since it had not yet been understood.  It is
interesting to note, however, that the proper light was shed on his theory
of electron and anti-electron symmetry (today climaxing in its
application to neutrinos and anti-neutrinos) and on his resulting
ability to eliminate the hypothesis known as the ``Dirac sea," a hypothesis
that was defined as ``extremely artificial and unsatisfactory,"
despite the fact that in general it had been uncritically accepted.

\h     The details of Majorana and Fermi's first meeting were
narrated by E. Segr\'e \cite{SE}: \ ``The first important work written
by Fermi in Rome [{`Su alcune propriet\`a statistiche dell'atomo'} (On certain
statistical properties of the atom)] is today known as the Thomas-Fermi
method\dots . When Fermi found that he needed the solution to a non-linear
differential equation characterized by unusual boundary conditions in order
to proceed, in a week of assiduous work with his usual energy, he
calculated the solution with a little hand calculator.  Majorana, who
had entered the Institute just a short time earlier and who was always
very skeptical, decided that Fermi's numeric solution probably was wrong
and that it would have been better to verify it.  He went home,
transformed Fermi's original equation into a Riccati equation, and
resolved it without the aid of any calculator, utilizing his extraordinary
aptitude for numeric calculation. When he returned to the Institute and
skeptically compared the little piece of paper on which he had written his
results to Fermi's notebook, and found that their results coincided
exactly, he could not hide his amazement." \  We have indulged in the
foregoing anecdote since the pages on which Majorana solved Fermi's
differential equation have in the end been found, and it has been shown
recently \cite{SalEsp} that he actually followed two independent (and
quite original) paths to the same mathematical result, one of them leading
to an Abel, rather than a Riccati, equation.

\h Majorana delivered his lectures only during the beginning of
1938, starting on Jan.13 and ending with his disappearance (March
26). But his activity was intense, and his interest for teaching
extremely high. For the benefit of his beloved students, and
perhaps also for writing down a book, he prepared careful notes
for his lectures. And ten of such lectures appeared in print in
1987 (see ref.\cite{Biblio}): and arose the admired comments of
many (especially British) scholars. The remaining six
lecture-notes, which had gone lost, have been rediscovered in 2005
by S.Esposito and A.Drago, and just appeared in
print\cite{biblio2}.

\section{Ettore Majorana's Published Papers.}

\h     Majorana published few scientific articles: nine, actually, besides
his sociology paper entitled ``Il valore delle leggi statistiche
nella fisica e nelle scienze sociali" (The value of statistical laws in
physics and the social sciences), which was however published not by
Majorana but (posthumously) by G. Gentile Jr., in {\it Scientia}
[{\bf 36} (1942) 55-56]. We already know that
Majorana switched from engineering to physics in 1928 (the year in
which he published his first article, written in collaboration with his
friend Gentile) and then went on to publish his works in theoretical
physics only for a very few years, practically only until 1933. \
Nevertheless, even his {\em published} works are a mine of ideas and
techniques of theoretical physics that still remains partially unexplored. \
Let us list his nine published articles:

\

\begin{enumerate}
\item[{(1)}]
  ``Sullo sdoppiamento dei termini Roentgen ottici a causa
dell'elettrone rotante e sulla intensit\`a delle righe del Cesio,"
in collaboration with Giovanni Gentile Jr., {\em Rendiconti Accademia
Lincei} {\bf 8} (1928) 229-233.

\item[{(2)}] ``Sulla formazione dello ione molecolare di He," {\em Nuovo Cimento}
{\bf 8} (1931) 22-28.

\item[{(3)}] ``I presunti termini anomali dell'Elio," {\em Nuovo Cimento}
{\bf 8} (1931) 78-83.

\item[{(4)}] ``Reazione pseudopolare fra atomi di Idrogeno," {\em Rendiconti
Accademia Lincei} {\bf 13} (1931) 58-61.

\item[{(5)}] ``Teoria dei tripletti {\em P'} incompleti,"
{\em Nuovo Cimento} {\bf 8} (1931) 107-113.

\item[{(6)}] ``Atomi orientati in campo magnetico variabile,"
{\em Nuovo Cimento} {\bf 9} (1932) 43-50.

\item[{(7)}] ``Teoria relativistica di particelle con momento intrinseco
arbitrario," {\em Nuovo Cimento} {\bf 9} (1932) 335-344.

\item[{(8)}] ``\"Uber die Kerntheorie," {\em Zeitschrift f\"ur Physik }
{\bf 82} (1933) 137-145; and ``Sulla teoria dei nuclei," {\em La Ricerca
Scientifica} {\bf 4}(1) (1933) 559-565.

\item[{(9)}] ``Teoria simmetrica dell'elettrone e del positrone," {\em Nuovo
Cimento} {\bf 14} (1937) 171-184.
\end{enumerate}

\

The first papers, written
between 1928 and 1931, concern atomic and molecular physics: mainly
questions of atomic spectroscopy or chemical bonds (within quantum mechanics,
of course).  As E. Amaldi has written \cite{Amaldi}, an in-depth examination
of these works leaves one struck by their superb quality: They reveal both a
deep knowledge of the experimental data, even in the minutest
detail, and an uncommon ease, without equal at that time, in the use
of the symmetry properties of the quantum states in order to
qualitatively simplify problems and choose the most suitable
method for their quantitative resolution.  Among the first papers,
``Atomi orientati in campo magnetico variabile" (Atoms oriented in
a variable magnetic field) deserves special mention.
It is in this article, famous among atomic physicists, that the effect
now known as the {\it Majorana-Brossel effect}
is introduced.  In it, Majorana predicts and calculates the modification
of the spectral line shape due to an oscillating magnetic field.  This
work has also remained a classic in the treatment of non-adiabatic
spin-flip.  Its results ---once generalized, as suggested by Majorana
himself, by Rabi in 1937 and by Bloch and Rabi in 1945--- established the
theoretical basis for the experimental method used to reverse the spin also
of neutrons by a radio-frequency field, a method that is still practiced
today, for example, in all polarized-neutron spectrometers. This Majorana
paper does moreover introduce the so-called {\it Majorana sphere} (to
represent complex spinors by a set of three, or two, points on the surface
of a sphere), as noted not long ago by R.Penrose\cite{Penrose} and
others\cite{Licata}.

\h Majorana's last three articles are all of such importance that none
of them can be set aside without comment.

\

\h      The article ``Teoria relativistica di particelle con
momento intrinseco arbitrario"  (Relativistic theory of particles
with arbitrary spin) is a typical example of a work that is so far
ahead of its time that it became understood and evaluated in depth
only many years later. Around 1932 it was commonly thought that
one could write relativistic quantum equations only in the case of
particles with zero or half spin.  Convinced of the contrary,
Majorana  ---as we know from his manuscripts--- began constructing
suitable quantum-relativistic equations \cite{BMR} for higher spin
values (one, three-halves, etc.); and he even devised a method for
writing the equation for a generic spin-value. But still he
published nothing, until he discovered that one could write a
single equation to cover  an infinite series of cases, that is, an
entire infinite family of particles of arbitrary spin (even if at
that time the known particles could be counted on one hand).  In
order to implement his programme with these ``infinite components"
equations, Majorana invented a technique for the representation of
a group several years before Eugene Wigner did. And, what is more,
Majorana obtained the infinite-dimensional unitary representations
of the Lorentz group that will be re-discovered by Wigner in his
1939 and 1948 works. The entire theory was re-invented by Soviet
mathematicians (in particular Gelfand and collaborators) in a
series of articles from 1948 to 1958 and finally applied by
physicists years later.  Sadly, Majorana's initial article
remained in the shadows for a good 34 years until D. Fradkin,
informed by E. Amaldi, released [{\em Am. J. Phys.} {\bf 34}
(1966) 314] what Majorana many years earlier had accomplished.

\

\h As soon as the news of the Joliot-Curie experiments reached Rome
at the beginning of 1932, Majorana understood that they had discovered the
``neutral proton" without having realized it.  Thus, even before the
official announcement of the discovery of the neutron, made soon
afterwards by Chadwick, Majorana was able to explain the structure and
stability of atomic nuclei with the help of protons and neutrons, antedating
in this way also the pioneering work of D. Ivanenko, as both Segr\'e and
Amaldi have recounted.  Majorana's colleagues remember that even before Easter
he had concluded that protons and neutrons (indistinguishable with respect
to the nuclear interaction) were bound by the ``exchange forces" originating
from the exchange of their spatial positions alone (and not also of their
spins, as Heisenberg would propose), so as to produce the alpha
particle (and not the deuteron) saturated with respect to the binding energy.
Only after Heisenberg had published his own article on the same
problem was Fermi able to persuade Majorana to meet his famous colleague
in Leipzig; and finally Heisenberg was able to convince Majorana to publish
his results in the paper ``\"Uber die Kerntheorie."
Majorana's paper on the stability of nuclei was immediately
recognized by the scientific community --a rare event, as we know, from
his writings-- thanks to that timely ``propaganda"  made by Heisenberg
himself.   \ We seize the present opportunity to quote two brief
passages from Majorana's letters from Leipzig. On February 14, 1933,
he writes his mother (the italics are
ours): ``The environment of the physics institute is very
nice. I have good relations with Heisenberg, with Hund, and with everyone
else. {\em I am writing some articles in German.  The first one is already
ready}...." The work that is already ready is, naturally, the
cited one on nuclear forces, which, however, remained {\em the only
paper} in German. Again, in a letter dated February 18, he tells
his father (we italicize): ``{\em I will publish in German, after having
extended it, also my
latest article which appeared in Nuovo Cimento}."  Actually, Majorana
published nothing more, either in Germany  or after his return to Italy,
except for the article (in 1937) of which we are about to speak. It is
therefore of importance to know that Majorana was  engaged in
writing other papers: in particular, he was expanding his article about
the infinite-components equations.

\

\h  As we said, from the existing manuscripts it appears that
Majorana was also formulating the essential lines of his symmetric theory
of electrons and anti-electrons during the years 1932-1933, even though
he published this theory only years later, when participating
in the forementioned competition for a professorship, under
the title ``Teoria simmetrica
dell'elettrone e del positrone" (Symmetrical theory of the electron and
positron), a publication that was
initially noted almost exclusively for having introduced the Majorana
representation of the Dirac matrices in real form. A consequence of this
theory is that a neutral fermion has to be identical with its
anti-particle, and
Majorana suggested that neutrinos could be particles of this type. As with
Majorana's other writings, this article also started to
gain prominence only decades later, beginning in 1957;  and nowadays expressions
like Majorana spinors,
Majorana mass, and Majorana neutrinos are fashionable.  As already
mentioned, Majorana's publications (still little known, despite it all)
is a potential gold-mine for physics.  Many years ago, for example,
Bruno Touschek noticed\cite{BrunoT} that the present article implicitly
contains also what he called the theory of the ``Majorana oscillator",
described by the simple equation \ $q+\omega^2q=\varepsilon \ \delta (t)$, \
where $\varepsilon$ is a constant and $\delta$ the Dirac
function\cite{Page70}. According to Touschek, the properties of the
Majorana oscillator are very interesting, especially in connection
with its energy spectrum: But no literature seems to exist yet
on it. \ Another example: much more recently,
C.Becchi pointed out how, in the first pages of
the present paper, a clear formulation of the
quantum action principle appears, the same principle that in later years,
through Schwinger's and Symanzik's works, for example, has brought about
quite important advances in quantum field theory.

\section{Ettore Majorana's Unpublished Papers.}

\h Majorana also left us several unpublished scientific
manuscripts, all of  which have been catalogued\cite{Catalog} and
kept at the ``Domus Galilaeana" of Pisa, Italy. Our analysis of
these manuscripts has allowed us to ascertain that all the
existing material seems to have been written by 1933; even the
rough copy of his last article, which Majorana proceeded to
publish in 1937
---as already mentioned--- seems to have been ready by 1933, the
year in which the discovery of the positron was confirmed.
Indeed, we are unaware of what he did in the following years from
1934 to 1938, except for a series of 34 letters written by
Majorana between March 17, 1931, and November 16, 1937, in reply
to his uncle Quirino ---a renowned experimental physicist and at a
time president of the Italian Physical Society--- who had been
pressing Majorana for theoretical explanations of his own
experiments.  By contrast, his sister Maria recalled that, even in
those years, Majorana
---who had reduced his visits to Fermi's Institute, starting from the beginning
of 1934 (that is, after his return from Leipzig)--- continued to study and
work at home many hours during the day and at night. \ Did he continue to
dedicate himself to physics? \  From a letter of his to Quirino, dated
January 16, 1936, we find a first answer, because we get to learn that
Majorana had been occupied ``since some time, with quantum electrodynamics";
knowing
Majorana's modesty and love for understatements, this no doubt means that
by 1935 Majorana had profoundly dedicated himself to original research in the
field of quantum electrodynamics.

\

\h Do any other unpublished scientific manuscripts of Majorana exist?  The
question, raised by his letters from Leipzig to his family, becomes of greater
importance when one reads also his letters addressed to the National Research
Council of Italy (CNR) during that period. In the first one (dated
January 21, 1933), Majorana asserts: ``At the moment,
I am occupied with the elaboration of a theory for the description of
arbitrary-spin particles that I began in Italy and of which I gave a
summary notice in {\it Nuovo Cimento}...."  In the second one (dated March 3,
1933) he even declares, referring to the same work: ``I have sent an article
on nuclear theory to Zeitschrift f\"ur Physik. I have the manuscript
of a new theory on elementary particles ready, and will send it to the
same journal in a few days."  Considering that the
article described here as a ``summary notice" of a new theory was
already of a very high level, one can imagine how interesting it would be
to discover a copy of its final version, which went unpublished.
[Is it still, perhaps, in the {\em Zeitschrift f\"ur Physik} archives?
Our own search ended in failure]. \
One must moreover not forget that the above-cited letter to Quirino Majorana, dated
January 16, 1936, revealed that his nephew continued to work on
theoretical physics even subsequently, occupying himself in depth, at
least, with quantum electrodynamics.

\h Some of Majorana's other ideas, when they did not remain concealed
in his own mind, have survived in the memories of his
colleagues. One such reminiscence we
owe to Gian Carlo Wick.  Writing from Pisa on October
16, 1978, he recalls: \  ``...The scientific contact [between Ettore and me],
mentioned by Segr\'e,
happened in Rome on the occasion of the `A. Volta Congress'
(long before Majorana's sojourn in Leipzig).  The conversation took
place in Heitler's company at a restaurant, and therefore without a
blackboard...; but even in the absence of details, what Majorana
described in words was a `relativistic theory of charged particles of
zero spin based on the idea of field quantization' (second quantization).
When much later I saw Pauli and Weisskopf's article [{\it Helv. Phys. Acta}
{\bf 7} (1934) 709], I remained absolutely convinced that what Majorana had
discussed was the same thing...."

\h We attach to this paper a short Bibliography.  Far from being exhaustive,
it provides only some references about the topics touched upon in this
paper.

\

\

\section{Acknowledgments.}

The author is quite grateful to all the Organizers of the Catania
Conference: A.Agodi, P.Castorina, F.Catara, E.Migneco, S.Lo Nigro,
F.Porto, E.Rimini, and in particular A.Rapisarda. \ Moreover, for
constant encouragement or collaboration, he thanks Franco Bassani,
Salvatore Esposito, Ignazio Licata, Fabio Majorana, Bruno Preziosi
and Silvano Sgrignoli.


\begin{thebibliography}{99}

\bibitem{Biblio}
{\em Ettore Majorana -- Lezioni all'Universit\`a di Napoli}, edited by
B.Preziosi (Bibliopolis pub.; Neaples, 1987).

\bibitem{biblio2}
{\em Lezioni di Fisica Teorica -- Ettore Majorana}, edited by
S.Esposito (Bibliopolis pub.; Neaples, 2006).

\bibitem{Amaldi}
E.Amaldi, {\em La Vita e l'Opera di E. Majorana} (Accademia dei Lincei; Roma,
1966); \ ``Ettore Majorana: Man and scientist," in {\em Strong
and Weak Interactions. Present problems}, edited by A.Zichichi
(Academic Press; New York, 1966); \ ``Ricordo di Ettore Majorana", in
{\em Giornale di Fisica} {\bf 9} (Bologna, 1968) p.300; \ E. Amaldi:
``From the discovery of the neutron to the discovery of
nuclear fission", in {\em Physics Reports} {\bf 111} (1984) pp.1--322; \
E. Amaldi: in {\em Il Nuovo Saggiatore} {\bf 4} (Bologna, 1988) p.13.

\bibitem{ER}
The {\em documents} that appear in this article (as well as all
the biographic documentation regarding E.Majorana, directly
discovered or collected by E.Recami in almost 30 years) can be
found in the book by E.Recami: {\em Il Caso Majorana: Epistolario,
Documenti, Testimonianze}, 2nd edition (Oscar, Mondadori pub.;
Milan, Italy, 1991), pp.230; and in particular in its 4th edition (Di
Renzo pub.; Rome, Italy, 2002: www.direnzo.it), pp.273. [Let us recall that
the material contained in that book is protected since 1986 by copyright in
favour of Maria Majorana together with E.Recami, and, now, with Di Renzo
publisher: and cannot be further reproduced without the written
permission of the right holders]. \ Of such a
book it exists a good translation into French, performed by
F. and Ph. Gueret (unpublished).

\bibitem{ER2}
See also E. Recami: ``I nuovi documenti sulla scomparsa di E.Majorana", in
{\em Scientia} 110 (1975) p.577; \ in {\em La Stampa} (Turin), 1st June and
29 June 1975; \ in {\em Corriere della Sera} (Milan), 19 October 1982 and
13 December 1983; \ ``Ricordo di Ettore Majorana a sessant'anni dalla sua
scomparsa: L'opera scientifica edita e inedita", in {\em Quaderni di Storia
della Fisica (S.I.F.)}, 5 (1999), pp.19-68; \ and moreover
VV.AA.: {\em Scienziati e tecnologi contemporanei: Enciclopedia
Biografica}, 3 vols., edited by E.Macorini (Mondadori pub.; Milan, 1974); \
M.Farinella: in {\em L'Ora} (Palermo), 22 and 23 July 1975; \ G.C.Graziosi:
``Le lettere
del mistero Majorana", in {\em Domenica del Corriere} (Milan), 28 November
1972; \ S.Ponz de Leon: ``Speciale News: Majorana", TV piece broadcast
on 30 September 1987 (``Canale Cinque"); \ B.Russo: ``Ettore Majorana -- Un
giorno di marzo", TV piece broadcast on 18 December 1990 (``RAI
Tre--Sicilia"); \  F.Tomandl e A.Stadler:
large radio piece on Ettore Majorana broadcast at the end of 2004
by the ``Austrian Radio", Vienna.

\bibitem{Kluwer} {\em Ettore Majorana - Notes on Theoretical Physics},
edited by S.Esposito, E.Majorana Jr., A. van der Merwe, and E.Recami
(Kluwer Acad. Pub.; Dordrecht, Boston and London, Nov. 2003).

\bibitem{Zanichelli} {\em Ettore Majorana -- Appunti Inediti di Fisica
Teorica}, edited by S.Esposito and E.Recami (Zanichelli pub.; Bologna, 2006).

\bibitem{BrunoT} B.Touschek: private communications (1976). \ Thanks are
due to E.Pessa for having forwarded to us, at that time, Touschek's comment.

\bibitem{Page70} See page 70 in the IV edition (2002) of the book by
Recami, quoted in ref.\cite{ER}.

\bibitem{BMR}
M.Baldo, R.Mignani e E.Recami: ``About a Dirac--like equation
for the photon, according to
Ettore Majorana'', {\em Lett. Nuovo Cimento} {\bf 11} (1974) 568 \
[interesting also for a possible physical interpretation of the photon
wave-function]. \ See also S.Esposito: {\em Found.
Phys.} {\bf 28} (1998) 231; \ and E.Giannetto; {\em Lett.
Nuovo Cimento}  {\bf 44} (1985) 140 e 145; \ ``Su alcuni manoscritti inediti
di E.Majorana", in {\em Atti IX Congresso Naz.le di Storia della Fisica},
edited by F.Bevilacqua (Milan, 1988) p.173.

\bibitem{SalEsp}
S.Esposito: ``Majorana solution of the Thomas-Fermi equation",
{\em Am. J. Phys.} {\bf 70} (2002) 852-856; \ ``Majorana transformation for
differential equations", {\em Int. J. Theor. Phys.} {\bf 41} (2002) 2417-2426.

\bibitem{Ponte}
B.Pontecorvo, {\em Fermi e la fisica moderna} (Editori Riuniti pub.; Rome,
1972); and in {\em Proceedings
International Conference on the History of Particle Physics,
Paris, July 1982}, {\em Physique} {\bf 43} (1982).

\bibitem{Zanic}
Cf. also, e.g., G.Enriques: {\em Via D'Azeglio 57} (Zanichelli pub.; Bologna,
1971).

\bibitem{SE}
E.Segr\'e: {\em Enrico Fermi, Fisico} (Zanichelli; Bologna (1971);  and
{\em Autobiografia di un Fisico} (Il Mulino pub.; Rome, 1995).

\bibitem{Penrose}
R.Penrose, ``Newton, quantum theory and reality," in {\em 300 Years of
Gravitation}, edited by  S.W.Hawking and W.Israel (University Press;
Cambridge, 1987); \ J. Zimba and R. Penrose, {\em Stud. Hist. Phil. Sci.}
{\bf 24} (1993) 697; \ R. Penrose: {\em Ombre della Mente (Shadows of the
Mind)} (Rizzoli pub.; 1996), pp.338--343 and 371--375; and the subsequent
studies, performed at Palermo, by C.Leonardi, F.Lillo, A.Vaglica e G.Vetri:
`` Quantum visibility,
phase-difference operators, and the Majorana Sphere", preprint (Phys.Dept.,
Univ. of Palermo, Italy; 1998); \ ``Majorana and Fano
alternatives to the Hilbert space", in {\em Mysteries, Puzzles, and Paradoxes
in Quantum Mechanics}, ed. by R.Bonifacio (A.I.P.; Woodbury, N.Y., 1999),
pp.312-315; \ F.Lillo: ``Aspetti Fondamentali nell'Interferometria
a Uno e Due Fotoni", PhD thesis (C.Leonardi supervisor), Dept. of Physics,
University of Palermo, 1998.

\bibitem{Licata}
{\em Majorana Legacy in Contemporary Physics}, edited by
I.Licata (Di Renzo pub.; Rome, 2006), which can be found also in the
electronic journal {\em EJTP Electr. J. Theor. Phys.} 3 (2006), issue no.10.

\bibitem{Catalog}
M.Baldo, R.Mignani, e E.Recami, ``Catalogo dei manoscritti scientifici
inediti di E. Majorana,"  in {\em Ettore Majorana -- Lezioni
all'Universit\`a di Napoli}, edited by B.Preziosi (Bibliopolis pub.;
Napoli, 1987), pp.175-197; \ and E. Recami, ``Ettore Majorana: L'opera
edita ed inedita," {\em Quaderni di Storia della Fisica (of the Giornale
di Fisica)} (S.I.F. pub.; Bologna) {\bf 5} (1999) 19-68.

\end{thebibliography}
\end{document}


\baselineskip 0.5cm

\vspace{1cm} \

\large{{\bf ETTORE MAJORANA:}}

\large {{\bf UN CENNO ALLE SUE OPERE (E VITA)}}

\vspace*{1 cm}

Erasmo Recami\\
{\em Facolt\`a di Ingegneria, Universit\`a statale di Bergamo, Bergamo, Italy;}\\
and {\em INFN-Sezione di Milano, Milan, Italy.}

\vspace{1cm}

{{\em Abstract  --- This article represents part of the invited talk
delivered by the author at the ``International Conference on
Majorana Legacy and the Physics of the XXI century" held at the
Department of Physics, University of Catania, Catania, Italy, on
5 and 6 October, 2006: Conference organized for celebrating
the 100th anniversary of the birth (in Catania) of the Italian
theoretical physicist Ettore Majorana; probably the brightest
Italian theoretician of the XX century (Enrico Fermi regarded
him as the brightest in the world of his time), even if to
some people Majorana is still known chiefly for his mysterious
disappearance, in 1938, when he was 31.  In this writing we
outline the significance of his main publications, as well as
his life: The biographical data being based on letters,
documents, testimonies discovered or collected by the author during
almost 30 years, and contained in the book by E.Recami, ``Il Caso
Majorana: Epistolario, Testimonianze, Documenti" (initially published,
in 1987 and 1991, by Mondadori, Milan, Italy, and presenly published,
in 2002, by Di Renzo Editore, Rome, Italy: www.direnzo.it).
At last, some information and comments are added with regard to the
scientific {\rm manuscripts} left unpublished by Majorana.}}

\

{Riassunto --- Questo articolo rappresenta la versione italiana di parte
del seminario su
invito tenuto dall'autore alla ``International Conference on Majorana's
Legacy and the Physics of the XXI Century" che ha avuto luogo il 5 e 6
ottobre 2006 presso il Dip.to di Fisica della Universit\`a di Catania:
Conferenza organizzata per celebrare il centenario della nascita (a
Catania) del fisico teorico Ettore Majorana; probabilmente il pi\`u
geniale teorico italiano del XX secolo (Enrico Fermi lo consider\`o
il pi\`u brillante del suo tempo nel mondo), anche se a qualcuno
Majorana e` ancora conosciuto soprattutto per la sua misteriosa
scomparsa, avvenuta nel 1938, quando aveva 31 anni. In questo scritto
descriviamo in sintesi tanto il significato delle sue pi\`u importanti
pubblicazioni, quanto la sua vita: i dati biografici essendo basati
su lettere, documenti, testimonianze scoperti o raccolti dall'autore
durante quasi 30 anni, e contenuti nel libro di E.Recami: ``Il Caso
Majorana: Epistolario, Testimonianze, Documenti" (pubblicato
inizialmente da Mondadori, Milano, nel 1987 e 1991, e al presente
pubblicato, nel 2002, da Di Renzo Editore, Roma: www.direnzo.it).
Infine, si aggiungono informazioni e commenti circa i {\em manoscritti}
scientifici lasciati inediti dal Majorana. \ [Oss.: Molto del materiale 
che segue \`e protetto da copyright, e non pu\`o essere ulteriormente riprodotto 
senza il permesso scritto di Fabio Majorana et al., di E.Recami, e dell'editore
Di Renzo].}

\vspace*{2cm}

\

\noindent {\bf Introduzione storico-biografica}

\

Il 2006 \`e l'anno del centenario della nascita di Ettore Majorana, il
grande fisico nato a Catania il 5 agosto 1906. Diciamo subito che Ettore
Majorana \`e a molti noto soprattutto per il mistero che circonda la sua
scomparsa, avvenuta (sembra tra Palermo e Napoli) quando aveva 31 anni.  Ma
il Majorana merita ogni attenzione essendo stato probabilmente il maggior
fisico teorico italiano del secolo da poco concluso.

\h La fama di Ettore Majorana pu\`o essere facilmente appoggiata, infatti,
a molte testimonianze come la seguente, dovuta alla memore penna di Giuseppe
Cocconi. Invitato da Edoardo Amaldi\cite{Amaldi}, dal CERN gli scrive (18 luglio 1965):
\begin{quotation}
\footnotesize{$<<$...Nel gennaio 1938, appena laureato, mi fu offerto, essenzialmente da te, di venire
a Roma per sei mesi nell'Istituto di Fisica dell'Universit\`a come assistente
incaricato, ed una volta l\'{\i} ebbi la fortuna di unirmi a Fermi, Bernardini
(che aveva avuto una Cattedra a Camerino pochi mesi prima) ed Ageno (lui
pure giovane laureato), nella ricerca dei prodotti di disintegrazione dei
``mesoni'' {\em mu} (allora chiamati mesotroni ed anche yukoni) prodotti
dai raggi cosmici [...]. Ricordo che, mentre ero con Fermi nella nostra
officinetta arriv\`o la notizia della scomparsa da Napoli del Majorana.
Fermi si dette da fare telefonando da varie parti sinch\'e, dopo alcuni
giorni, si ebbe l'impressione che non lo si sarebbe ritrovato pi\'u.

Fu allora che Fermi, cercando di farmi capire che cosa significasse tale perdita, si espresse
in modo alquanto insolito, lui che era cos\'{\i} serenamente severo quando si trattava
di giudicare il prossimo. Ed a questo punto vorrei ripetere le sue parole, cos\'{\i}
come da allora me le sento risuonare nella memoria: {\em ``Perch\'e, vede, al mondo
ci sono varie categorie di scienziati; gente di secondo e terzo rango, che fan
del loro meglio ma non vanno molto lontano. C'\`e anche gente di primo rango, che
arriva a scoperte di grande importanza, fondamentali per lo sviluppo della scienza}
(e qui ho netta l'impressione che in quella categoria volesse mettere se stesso).
{\em Ma poi ci sono i geni, come Galileo e Newton. Ebbene, Ettore era uno di quelli.
Majorana aveva quel che nessun altro mondo ha...''}$>>$}.
\end{quotation}

\

\h Enrico Fermi [uno dei maggiori fisici della nostra epoca; per quello
che ha fatto nel 1942 a Chicago, con la costruzione della prima ``pila atomica'',
il suo nome diverr\`a forse leggendario come quello di Prometeo...]
si espresse in maniera per lui insolita anche in un'altra occasione, il 27 luglio 1938
(dopo la scomparsa di Majorana, avvenuta il sabato 26 marzo 1938), scrivendo
da Roma al primo ministro Mussolini onde chiedere una intensificazione delle
ricerche di Ettore:
\begin{quotation}
\footnotesize{
$<<$ Io non esito a dichiararVi, e non lo dico quale espressione
iperbolica, che fra tutti gli studiosi italiani e stranieri che ho avuto occasione
di avvicinare il Majorana \`e fra tutti quello che per profondit\`a di ingegno
mi ha maggiormente colpito$>>$.}
\end{quotation}

\

\h E un testimone diretto, Bruno Pontecorvo\cite{Ponte}, aggiunge: $<<$Qualche
tempo dopo
l'ingresso nel gruppo di Fermi\cite{Zanic}, Majorana possedeva gi\`a una erudizione
tale ed aveva raggiunto un tale livello di comprensione della fisica da
potere parlare con Fermi di problemi scientifici da pari a pari. Lo stesso
Fermi lo riteneva il pi\'u grande fisico teorico dei nostri tempi. Spesso
ne rimaneva stupito [...]. Ricordo esattamente queste parole di Fermi:
``Se un problema \`e gi\`a posto, nessuno al mondo lo pu\`o
risolvere meglio di Majorana".$>>$

\h  Il mito della ``scomparsa", quindi,  ha contribuito a dare a Majorana
null'altro che la notoriet\`a che gli spettava, per essere un genio
e un genio molto avanzato rispetto ai suoi tempi.

\h Il Majorana, passato a fisica alla fine del '27, si
laure\`o con Fermi il 6 luglio 1929, e continu\`o a collaborare col famoso
gruppo di Enrico Fermi e Franco Rasetti (nato per volont\`a e attiva opera di
Orso Mario
Corbino): i cui fisici teorici ---in ordine di ingresso nel gruppo---
furono Ettore Majorana, Gian Carlo Wick, Giulio Racah, Giovanni Gentile jr.,
Ugo Fano, Bruno Ferretti, e Piero Caldirola.  Membri del sottogruppo
sperimentale furono Emilio Segr\'e, Edoardo Amaldi, Bruno Pontecorvo,
Eugenio Fubini, Mario Ageno, Giuseppe Cocconi, oltre all'ottimo chimico
Oscar D'Agostino. Successivamente, Majorana consegu\'{\i} la Libera Docenza
in fisica teorica il 12 novembre 1932; trascorse circa sei mesi a Lipsia con
Werner Heisenberg durante il 1933; e quindi, per ragioni ignote, interruppe
la sua frequentazione del gruppo dei ``ragazzi di via
Panisperna".  Smise perfino di pubblicare i risultati delle proprie ricerche
(che gi\`a in precedenza aveva drasticamente selezionato basandosi sui
suoi eccezionali
spirito critico e amore per il rigore e le vere innovazioni); a parte
l'articolo ``Teoria simmetrica dell'elettrone e del positrone," gi\`a pronto
fin dal 1933, e che, stimolato dai suoi colleghi, Majorana tir\`o fuori da un
cassetto e pubblic\`o in occasione del Concorso nazionale del 1937 a tre posti
di professore ordinario di fisica teorica.

\h In relazione a quest'ultimo punto, ricordiamo che nel 1937 i concorrenti
furono numerosi, e molti di essi di elevato valore; soprattutto quattro:
Ettore Majorana, Giulio Racah (ebreo, che successivamente passer\`a
da Firenze in Israele fondandovi la fisica teorica), GianCarlo Wick
(di madre torinese e nota antifascista), e Giovanni Gentile jr.
(figlio dell'omonimo filosofo, gi\`a ministro ---come si direbbe
ora--- della Pubblica Istruzione, e ideatore delle ``parastatistiche'' in
meccanica quantica). La  commissione giudicatrice era costituita da:
Enrico Fermi (presidente), Antonio Carrelli, Orazio Lazzarino, Enrico Persico
e Giovanni Polvani. Su raccomandazione della commissione giudicante, il
ministro dell'Educazione Nazionale Giuseppe Bottai nomin\`o il Majorana
professore di fisica teorica all'Universit\`a di Napoli per la sua
``grande e meritata fama", al di fuori del Concorso stesso. La
Commissione, invero, aveva dichiarato per iscritto al Ministro di esitare ad
applicare a lui le normali procedure concorsuali; allegando il seguente
giudizio\cite{ER,ER2}:
\begin{quotation}
\footnotesize{$<<$...Senza elencarne i lavori, tutti notevolissimi
per l'originalit\`a dei metodi impiegati e per l'importanza dei risultati
raggiunti, ci si limita qui alle seguenti segnalazioni:

Nelle teorie nucleari moderne il contributo portato da questo ricercatore con la
introduzione
delle forze dette ``Forze di Majorana'' \`e universalmente riconosciuto, tra i
pi\'u fondamentali, come quello che permette di comprendere teoricamente le ragioni
della stabilit\`a dei nuclei. I lavori del Majorana servono oggi di base alle pi\'u
importanti ricerche in questo campo.

Nell'atomistica spetta al Majorana il merito di aver risolto, con semplici ed eleganti
considerazioni  di simmetria, alcune tra le pi\'u intricate questioni sulla
struttura degli spettri.

In un recente lavoro infine ha escogitato un brillante metodo che permette di
trattare in modo simmetrico l'elettrone positivo e negativo, eliminando finalmente
la necessit\`a di ricorrere all'ipotesi estremamente artificiosa ed insoddisfacente di una
carica elettrica infinitamente grande diffusa in tutto lo spazio,
questione che era stata invano affrontata da molti
altri studiosi$>>$}.
\end{quotation}

\h Uno dei lavori pi\'u importanti di Ettore, quello in cui introduce la sua
``equazione a infinite componenti'' (di cui diciamo in seguito), non \`e
menzionato: ancora non era stato capito.
\`E interessante notare, per\`o, che viene dato giusto rilievo alla sua teoria
simmetrica per l'elettrone e l'anti-elettrone (oggi in auge, per la sua
applicazione a neutrini e anti-neutrini); e a causa
della capacit\`a di eliminare l'ipotesi cosiddetta ``del {\em mare} di Dirac''
[P.A.M. Dirac, premio Nobel 1933]: ipotesi che viene definita ``estremamente
artificiosa e insoddisfacente'', nonostante che essa dai pi\'u sia sempre stata
accettata in maniera acritica.

\

\h I dettagli del primo incontro di Majorana con Fermi ci illuminano
circa alcuni aspetti, scientifici e no, di Ettore. Essi sono noti da
quando li ha narrati Segr\'e\cite{Segre}; ma vale la pena di rileggerli con
attenzione: $<<$Il primo lavoro
importante scritto da Fermi a Roma [{\em su alcune propriet\`a
statistiche dell'atomo\/}]$\ldots$ \`e oggi noto come metodo di
Thomas-Fermi $\ldots$Quando Fermi trov\`o che per procedere gli
occorreva la soluzione di un'equazione differenziale non lineare,
caratterizzata da condizioni al contorno insolite, con la sua
abituale energia in una settimana di assiduo lavoro calcol\`o la
soluzione con una piccola calcolatrice a mano. Majorana, che era
entrato da poco in Istituto e che era sempre molto scettico, decise
che probabilmente la soluzione numerica di Fermi era sbagliata e che
sarebbe stato meglio verificarla. And\`o a casa, trasform\`o durante la serata
e la notte l'equazione originale di Fermi in una equazione del tipo di Riccati e
la risolse senza l'aiuto di nessuna calcolatrice, servendosi della
sua straordinaria attitudine al calcolo numerico... Quando il mattino dopo
torn\`o in Istituto confront\`o con aria
scettica il pezzetto di carta, su cui aveva riportato i dati
ottenuti, col quaderno di Fermi, e quando trov\`o che i risultati
coincidevano esattamente non pot\'e nascondere la sua meraviglia$>>$.

\h Abbiamo indugiato sul precedente aneddoto dato che le pagine con la
soluzione in forma chiusa trovata dal Majorana per l'equazione differenziale
di Fermi ---equazione che Fermi, ripetiamolo, non era riuscito a risolvere
analiticamente--- sono state da noi alfine scoperte: e si \`e cos\`{\i} potuto
recentemente mostrare\cite{Espos} che Majorana segu\`{\i} in realt\`a due indipendenti
metodi (molto originali) per giungere ai medesimi risultati, uno
dei quali lo condusse ad una equazione di Abel, piuttosto che di Riccati.
Il secondo cammino costituisce una novit\`a anche per la matematica attuale.
La comprensione dettagliata di quanto fatto da Majorana in quelle poche ore
ha richiesto al fisico teorico S.Esposito, di Napoli, circa due mesi di intensa
applicazione... [sfociata
comunque in due pubblicazioni, la prima delle quali sull'{\em American
Journal of Physics\/}].

\

\

\noindent{\bf Gli articoli {\em pubblicati} da Ettore Majorana}

\

Ettore scrisse pochi articoli scientifici: nove; oltre allo scritto
semi-divulgativo ``Il valore delle leggi statistiche nella fisica e
nelle scienze sociali'', pubblicato postumo su {\em Scientia}
[{\bf 36} (1942) 55-66] a cura di G. Gentile. Si ricordi che
Majorana pass\`o da ingegneria a fisica alla fine del 1927 o
forse agli inizi del 1928 (anno in cui
pubblic\`o gi\`a un articolo, il primo: scritto insieme con l'amico
Gentile), e poi si dedic\`o alla pubblicistica in fisica teorica solo per
pochissimi anni, in pratica fino al 1933. Ciononostante, anche i soli
lavori da lui {\em publlicati} sono una miniera di idee e tecniche di
fisica teorica che rimane tuttora parzialmente inesplorata: per fortuna
quest'anno la SIF riunir\`a in un unico volume tutte le pubblicazioni
del Majorana, con la loro traduzione in inglese].

Elenchiamo i suoi nove articoli pubblicati:

\

\begin{enumerate}
\item[{(1)}]
  ``Sullo sdoppiamento dei termini Roentgen ottici a causa
dell'elettrone rotante e sulla intensit\`a delle righe del Cesio,"
in collaboration with Giovanni Gentile Jr., {\em Rendiconti Accademia
Lincei} {\bf 8} (1928) 229-233.

\item[{(2)}] ``Sulla formazione dello ione molecolare di He," {\em Nuovo Cimento}
{\bf 8} (1931) 22-28.

\item[{(3)}] ``I presunti termini anomali dell'Elio," {\em Nuovo Cimento}
{\bf 8} (1931) 78-83.

\item[{(4)}] ``Reazione pseudopolare fra atomi di Idrogeno," {\em Rendiconti
Accademia Lincei} {\bf 13} (1931) 58-61.

\item[{(5)}] ``Teoria dei tripletti {\em P'} incompleti,"
{\em Nuovo Cimento} {\bf 8} (1931) 107-113.

\item[{(6)}] ``Atomi orientati in campo magnetico variabile,"
{\em Nuovo Cimento} {\bf 9} (1932) 43-50.

\item[{(7)}] ``Teoria relativistica di particelle con momento intrinseco
arbitrario," {\em Nuovo Cimento} {\bf 9} (1932) 335-344.

\item[{(8)}] ``\"Uber die Kerntheorie," {\em Zeitschrift f\"ur Physik }
{\bf 82} (1933) 137-145; and ``Sulla teoria dei nuclei," {\em La Ricerca
Scientifica} {\bf 4}(1) (1933) 559-565.

\item[{(9)}] ``Teoria simmetrica dell'elettrone e del positrone," {\em Nuovo
Cimento} {\bf 14} (1937) 171-184.
\end{enumerate}

\

Esaminiamo brevemente i detti articoli. I primi, redatti tra il 1928 e il
1931, riguardano problemi di fisica atomica e molecolare: per lo
pi\'u questioni di spettroscopia atomica o di legame chimico (sempre,
s'intende, nell'ambito della meccanica quantistica). Come scrive E.
Amaldi, un esame approfondito di questi lavori lascia colpiti per la
loro alta classe: essi rivelano sia una profonda conoscenza dei dati
sperimentali anche nei pi\'u minuti dettagli, sia una disinvoltura
non comune, soprattutto a quell'epoca, nello sfruttare le propriet\`a
di simmetria degli ``stati quantistici'' per semplificare
qualitativamente i problemi e per scegliere la via pi\'u opportuna
per la risoluzione quantitativa.  Tra questi primi articoli ne
scegliamo un solo:

\h ``Atomi orientati in campo magnetico variabile'' apparso sulla
rivista {\em Nuovo Cimento}, vol.{\bf 9} (1932) pp.43-50. \`E
l'articolo, famoso tra i fisici atomici, in cui viene introdotto
l'effetto ora noto come Effetto Majorana-Brossel. In esso Ettore
prevede e calcola la modificazione della forma delle righe spettrali
dovuta a un campo magnetico oscillante; e ci\`o in connessione a un
esperimento tentato a Firenze qualche anno prima (bench\'e senza
successo) da G. Bernardini ed E. Fermi. Questo lavoro \`e rimasto
anche un classico della trattazione dei processi di ribaltamento
dello {\em spin} (o ``spin-flip''). I suoi
risultati ---una volta estesi, come suggerito dallo stesso Majorana,
da Rabi nel 1937 e quindi, nel 1945, da Bloch e Rabi (i quali,
entrambi premi Nobel [Rabi: 1944; Bloch: 1952], contribuirono a
diffondere  quanto trovato da Ettore tredici anni prima)--- hanno
costituito la base teorica del metodo sperimentale usato per
ribaltare anche lo {\em spin} dei neutroni con un campo a
radiofrequenza: metodo impiegato ancor oggi, ad esempio, in tutti gli
spettrometri a neutroni polarizzati. \ In questo articolo viene introdotta
anche la cosiddetta ``Sfera di
Majorana" (per rappresentare spinori mediante set di punti di una
superficie sferica), di cui ha parlato entusiasticamente ---per esempio---
Roger Penrose\cite{Penrose}, anche nei suoi ultimi libri semi-divulgativi
(si vedano in
Bibliografia le citazioni di Penrose, e Zimba \& Penrose, quelle pi\'u
recenti di C.Leonardi et al., e il libro uscito nel 2006 a cura di
I.Licata\cite{Licata}): Majorana, infatti, si era accorto, tra
l'altro, del fatto che gli spinori sono invarianti per rotazioni spaziali.

\h Gli ultimi tre articolo di Ettore sono tutti di tale importanza che
nessuno di essi pu\`o restare senza commento.

\subsection*{5.4. L'equazione a infinite componenti}

\paragraph*{}
L'articolo ``Teoria relativistica di particelle con momento
intrinseco arbitrario'' {\em Nuovo Cimento}, vol.{\bf 9} (1932) pp.335-344)
\`e il tipico esempio di lavoro che precorre talmente i
tempi da venire compreso e valutato a fondo solo molti anni dopo.

\h A quel tempo era opinione comune\cite{BMR} che si potessero scrivere
equazioni quantistiche compatibili con la Relativit\`a (cio\`e
``relativisticamente invarianti'') solo nel caso di particelle a spin
zero o un mezzo. Convinto del contrario, Ettore comincia a costruire
opportune equazioni quanto-relativistiche per i successivi valori
possibili per lo spin (uno, tre mezzi, ecc.), arrivando a dare le regole
anche per la costruzione di tale equazione per un valore generico dello
spin;  finch\'e scopre che si
pu\`o scrivere un'{\em unica} equazione rappresentante una serie
infinita di casi, cio\`e un'intera famiglia infinita di particelle a
spin qualsiasi (si ricordi che allora le particelle note ---che ora
sono centinaia--- si contavano sulle dita di una mano!). Tralascia
allora tutti i singoli casi studiati ---senza pi\'u pubblicarli--- e
si dedica solo a queste equazioni ``a infinite componenti'', senza
trascurare l'osservazione che esse possono descrivere non solo
particelle ordinarie ma anche tachioni.

\h Per realizzare questo programma inventa una tecnica per la
``rappresentazione di un gruppo'' vari anni prima della ``scoperta''
di queste tecniche  da parte di Eugene Wigner (premio Nobel 1963).
Pi\'u ancora, Majorana ricorre per la prima volta ---inventandole---
alle rappresentazioni unitarie del Gruppo di Lorentz {\em a infinite
dimensioni}: rappresentazioni riscoperte da Wigner in lavori del 1939
e 1948. Per comprendere l'importanza di quest'ultimo aspetto,
rifacciamoci a quanto Ettore stesso ---pur tanto schivo--- riferisce
a suo padre da Lipsia il 18 febbraio 1933: \ $<<$Nell'ultimo mio
articolo apparso sul ``Nuovo Cimento'' \`e contenuta una importante
scoperta matematica, come ho potuto accertarmi mediante un colloquio
col professor van der Waerden, olandese che insegna qui, una delle
maggiori autorit\`a in teoria dei gruppi$>>$.

\h Questa teoria \`e stata reinventata da matematici sovietici (in
particolare Gelfand e collaboratori) in una serie di articoli del
1948-1958, e finalmente applicata dai fisici in anni ancora pi\'u
tardi. L'articolo iniziale di Ettore, anzi, rimarr\`a in ombra per
ben 34 anni, cio\`e fino a quando Amaldi lo traduce e segnala al
fisico americano D.Fradkin, il quale a sua volta strabilia i teorici
delle alte energie rendendo finalmente di pubblico dominio, nel
1966 [D. Fradkin: {\em American Journal of Physics} {\bf 34}
(1966) 314], quanto compiuto da Majorana tanti anni prima. Dalla data
del 1966, la fama di Ettore comincia a crescere costantemente anche
tra i fisici delle particelle fondamentali.

\subsection*{5.5. Le forze di scambio}

\paragraph*{}
Non appena, al sorgere del 1932, giunge a Roma notizia degli
esperimenti dei Joliot-Curie [premi Nobel 1935 per la chimica],
Ettore comprende che essi avevano scoperto il ``protone neutro''
senza accorgersene [cos\`{\i} si espresse il Majorana, anche se da
parecchi anni era gi\`a usato il termine neutrone, la cui esistenza
era da molto tempo ventilata]. Prima ancora, quindi, che ci fosse l'annuncio
ufficiale della scoperta del {\em neutrone}, effettuata il mese
successivo da
Chadwick [premio Nobel 1935 per la fisica], Majorana \`e in grado
di spiegare la struttura e la stabilit\`a dei nuclei atomici
mediante protoni e neutroni. (I suoi manoscritti inediti ci dicono
che egli si era gi\`a cimentato su questo problema ricorrendo,
invano, a protoni ed elettroni: che erano le uniche particelle in
precedenza note). Ettore precorse cos\'{\i} anche il lavoro
pionieristico di D.Ivanenko. Ma non volle pubblicarne nulla, n\'e
permise a Fermi di parlarne a Parigi agli inizi di luglio: ci\`o \`e
narrato da Segr\'e e da Amaldi. I suoi colleghi ricordano che gi\`a
prima di Pasqua era giunto alle conclusioni pi\'u importanti della
sua teoria: che protoni e neutroni fossero legati da forze
quantistiche originate semplicemente dalla loro {\em
indistinguibilit\`a}; cio\`e da ``forze di {\em scambio}'' delle
rispettive posizioni spaziali (e non anche degli spin, come invece
far\`a Heisenberg), cos\'{\i} da ottenere la particella alfa (e non
il deutone) quale sistema saturato rispetto alla  energia di legame.

\h Solo dopo che Heisenberg pubblica il proprio articolo sullo stesso
argomento, Fermi riesce a indurre Majorana a recarsi a Lipsia
presso il grande collega. E, finalmente, Heisenberg sa convincere
Ettore a pubblicare (anche se tanto in ritardo) i propri risultati:
``\"{U}ber die Kerntheorie'', lavoro apparso il 3 marzo 1933 su {\em
Zeitschrift f\"{u}r Physik}, vol.{\bf 82} (1933) pp.137-145.

\h Le forze ``di scambio'' nucleari furono chiamate forze di
Heisenberg-Majorana. Ettore ne parla al padre, con grande modestia,
nella stessa lettera prima citata (del 18.2.1933): \ $<<$Ho
scritto un articolo sulla struttura dei nuclei che a Heisenberg \`e
piaciuto molto bench\'e contenesse alcune correzioni a una sua
teoria$>>$. Sempre su questo lavoro scrive pochi giorni dopo, il 22
febbraio, alla madre: $<<$Nell'ultimo ``colloquio'', riunione
settimanale a cui partecipano un centinaio tra fisici, matematici,
chimici, etc., Heisenberg ha parlato della teoria dei nuclei e mi ha
fatto molta r\'eclame a proposito di un lavoro che ho scritto qui.
Siamo diventati abbastanza amici$\ldots >>$.

\h Probabilmente la pubblicazione sulla stabilit\`a dei nuclei venne
subito riconosciuta dalla comunit\`a scientifica (in particolare dai
fisici nucleari) ---evento raro, come sappiamo, per gli scritti di
Ettore--- anche grazie a questa opportuna ``propaganda'' fattane da
Heisenberg, che proprio pochi mesi dopo ricever\`a il premio Nobel.

\h L'avversione a pubblicare le proprie scoperte, quando esse fossero
risultate, all'esame del suo senso ipercritico, di carattere non
abbastanza generale o espresse in forma matematica non abbastanza
stringente ed elegante, divenne per Ettore anche motivo di vezzo.
Racconta Amaldi: \ $<<$Talvolta nel corso di una conversazione con
qualche collega
diceva quasi incidentalmente di aver fatto durante la sera precedente
il calcolo o la teoria di un fenomeno non chiaro che era caduto sotto
l'attenzione sua o di qualcuno di noi in quei giorni. Nella
discussione che seguiva, sempre molto laconica da parte sua, Ettore a
un certo punto tirava fuori dalla tasca il pacchetto delle sigarette
Macedonia (era un fumatore accanito) sul quale erano scritte, in una
calligrafia minuta ma ordinata, le formule principali della sua
teoria o una tabella di risultati numerici. Copiava sulla lavagna
parte dei risultati, quel tanto che era necessario per chiarire il
problema, e poi, finita la discussione e fumata l'ultima sigaretta,
accartocciava il pacchetto nella mano e lo buttava nel cestino$>>$.

\h Estremamente interessanti sono pure due altri passi di lettera. Il
14.2.1933, sempre da Lipsia, Majorana racconta alla madre: \ $<< \ldots$
L'ambiente dell'istituto fisico \`e molto simpatico. Sono in ottimi
rapporti con Heisenberg, con Hund e con tutti gli altri. {\em Sto
scrivendo alcuni articoli in tedesco. Il primo \`e gi\`a pronto}, e
spero di eliminare qualche confusione linguistica durante la
correzione delle bozze$>>$. \ Il lavoro ``gi\`a pronto'' \`e
naturalmente quello sulle forze nucleari di cui si sta parlando; il
quale, per\`o, rimase l'{\em unico} in lingua tedesca.

\h Ancora: nella lettera del 18 febbraio dichiara al padre. \ $<<\ldots$
{\em pubblicher\`o in tedesco, estendendolo, anche l'ultimo mio
articolo apparso sul ``Nuovo Cimento''}$>>$.

\h In realt\`a Ettore non pubblic\`o pi\'u nulla, n\'e in Germania, n\'e al
rientro in Italia, a parte  l'articolo (del 1937) di cui stiamo per dire. \
Di notevole importanza \`e quindi sapere che Ettore stesse scrivendo
altri lavori: in particolare, che stesse {\em estendendo} il suo
articolo sulla equazione a infinite componenti. Nel brano alla madre,
\`e probabile si riferisca pure alla sua teoria simmetrica di
elettrone e anti-elettrone, pubblicata solo quattro anni pi\'u tardi.

\subsection*{5.6. Il neutrino di Majorana}

\paragraph*{}
Dai manoscritti ritrovati pare, come si \`e detto, che Majorana
formulasse in quegli stessi anni (1932-33) le linee essenziali anche
della sua teoria  simmetrica per l'elettrone e l'anti-elettrone: che
le formulasse, cio\`e, non appena si diffuse la notizia della scoperta
dell'anti-elettrone, o ``positrone''. Anche se Ettore pubblica tale
teoria solo molto pi\'u tardi, accingendosi a partecipare al Concorso
a cattedra di cui sappiamo: ``Teoria simmetrica dell'elettrone e del
positrone'', {\em Nuovo Cimento}, vol.{\bf 14} (1937) pp.171-184.
Questa pubblicazione viene inizialmente notata quasi esclusivamente per aver
introdotto la famosa {\em rappresentazione di Majorana} delle
``matrici di Dirac'' in forma reale\footnote{Si noti, per\`o, che
l'algebra $I\!\!R (4)\simeq I\!\!R_{3,1}$ cos\'{\i} introdotta da
Majorana \`e del tutto diversa dall'algebra $C\!\!\!\!I\ (4)\simeq I\!\!R_{4,1}$
introdotta da Dirac. Osserviamo, en passant, che l'algebra di
Majorana \`e una delle {\em due} algebre associabili in maniera naturale
allo spazio di Minkowski (la seconda essendo $I\!\!R_{1,3}\simeq
I\!\!H(2)$, ove $I\!\!H(2)$ \`e l'algebra delle matrici
quaternioniche $2\times 2$).}.  Conseguenza di tale teoria \`e  che un
``fermione'' neutro debba coincidere con la propria antiparticella:
ed Ettore suggerisce che i neutrini possano essere particelle di
questo tipo.

\h Ettore ci teneva molto a questa sua elaborazione teorica; ci\`o \`e
testimoniato da Carrelli, che ne discusse con Ettore durante il breve
periodo di lezioni a Napoli.

\h Come per altri scritti di Majorana, anche questo articolo ha
cominciato ad avere fortuna solo vent'anni dopo, a partire dal 1957.
Dopo di che ha goduto di fama via via crescente tra i fisici delle
particelle relativistiche e delle teorie di campo\footnote{Nel 1981,
ad esempio, una rivista giapponese di fisica ha ripubblicato in
lingua inglese (con traduzione a cura di Luciano Maiani) questo
articolo di circa quarantacinque anni prima.}. Ora sono di gran moda
espressioni come ``spinori di Majorana'', ``massa di
Majorana'', ``neutrini di Majorana'' (e perfino  ``majoroni"). \ Le
pubblicazioni di Majorana (ancora poco note, nonostante tutto) sono per
la Fisica, lo si \`e detto, una continua fonte di ispirazione. Recentemente,
ad esempio, Carlo Becchi ha osservato
come nelle prime pagine di questo scritto si trovi una formulazione
estremamente chiara del principio d'azione quantistico, che in anni
successivi, attraverso i lavori di Schwinger e Symanzik, ha portato agli
sviluppi recenti pi\'u importanti di teoria dei campi quanto-relativistici. \

Segnaliamo inoltre che (attraverso E.Pessa) dal grande fisico Bruno
Touschek ci giunse a suo tempo la segnalazione che questo articolo contiene
pure la teoria\cite{ER} dell'``oscillatore di Majorana", fino ad oggi
pochissimo studiata.

\

\

\noindent{\bf I manoscritti inediti di Ettore Majorana}

\

Ma Ettore ci ha lasciato anche molti manoscritti scientifici
inediti, pure depositati presso la ``Domus Galilaeana'' (ora,
purtroppo, in situazione critica: come quasi tutte le istituzioni
culturali italiane) di cui
abbiamo redatto un catalogo\cite{Catalog} in collaborazione
con M.Baldo e R.Mignani. \ L'analisi di questi manoscritti permette
di rilevare: \ (i) come Ettore fosse estremamente diligente e preciso
nel lavoro. Tutte le sue scoperte risultano precedute da una
indefessa serie di calcoli, fatti e rifatti: anche per i pi\'u
dotati, naturalmente, la scienza non pu\`o essere solo un semplice
gioco di intuizioni, come invece la leggenda aveva voluto farci
credere; \ (ii) che, fra il materiale inedito, parecchi spunti hanno ancora
un interesse scientifico {\em attuale}: alcune centinaia di pagine possono
essere utili in maniera significativa per la ricerca contemporanea; ma solo
poche pagine sono state da noi finora interpretate e
pubblicate;\cite{BMR,Espos,Kluwer}  \ (iii) che tutto il materiale noto
{\em sembra} scritto entro il 1933 (anche la  bozza
dell'ultimo articolo, sulla ``Teoria simmetrica dell'elettrone e del
positrone'', che, come si \`e gi\`a detto, Ettore pubblicher\`a alle soglie
del concorso a
cattedra nel 1937, pare fosse gi\`a pronta dal 1933, anno in cui si
ha la conferma della scoperta ---appunto--- del positrone); \ (iv) che quasi
nulla ci \`e noto di ci\`o che egli fece negli anni a seguire (1934--1938). \
A parte una lunga serie di 34 lettere di risposta, scritte da Ettore in
quegli anni (precisamente dal 17.3.31
fino al 16.11.37) allo zio Quirino, il quale lo sollecitava a fornire una
spiegazione teorica dei risultati dei propri esperimenti.  Queste lettere
sono di carattere essenzialmente tecnico (lo zio Quirino era un fisico
sperimentale di grandissima abilit\`a, che aveva occupato anche il ruolo
di presidente della Societ\`a Italiana di Fisica) e confermano
che pure negli ultimi anni Ettore ben sapeva tornare alla fisica, sempre
con le sue doti di eccelso teorico [tali lettere ci sono state consegnate
dalla figlia di Quirino, Silvia Toniolo M., per interessamento di Franco
Bassani].

\h Invero la sorella Maria ricordava che anche in quegli anni Ettore ---il
quale aveva diradato sempre pi\'u le sue visite all'Istituto, a cominciare dalla
fine del 1933, cio\`e dal suo rientro da Lipsia--- continu\`o a studiare e
lavorare a casa parecchie ore al giorno; e la notte. Si diede Ettore solo a
studi di letteratura e filosofia (amava particolarmente Pirandello,
Schopenhauer e Shakespeare), o di ``teoria dei giochi" e strategia navale
(sua passione fin dall'infanzia), nonch\'e di economia, di politica e infine
di medicina; oppure continu\`o a dedicarsi anche alla Fisica?  Dalla lettera
a Quirino del 16.1.1936 ci viene una risposta; perch\'e veniamo a
sapere che Ettore si occupava
``da qualche tempo di elettrodinamica quantistica".  Conoscendo la modestia
di Ettore nell'esprimersi, ci\`o significa che durante l'anno 1935 Majorana
si era dedicato a fondo a ricerche originali nel settore  ---per lo meno---
della elettrodinamica quantistica. \ E ancora nel 1938, a Napoli, Carrelli
avr\`a l'impressione che Ettore stesse lavorando a qualcosa di importante,
di cui non voleva parlare. \ Ma lumi ancora pi\`u significativi ci sono giunti
dalle lettere inviate, da Lipsia, ai propri genitori, lettere che abbiamo
sopra citate, e, sempre da Lipsia, al C.N.R.: delle quali diremo. Per di
pi\`u, \`e stato recentemente scoperto che il Majorana, per gli anni
accademici 1933/34, 1934/35 e 1935/6, per ben tre volte, aveva chiesto
all'Istituto di Fisica di Roma di poter tenere dei corsi ``liberi" di
fisica teorica: cosa che gli era permessa dall'essere egli Libero Docente
(il direttore O.M.Corbino sempre approv\`o, ma pare che il Nostro non
tenne mai i desiderati corsi: probabilmente perch\'e troppo pochi studenti,
allora, si rendevano conto dell'importanza degli argomenti proposti dal
Majorana). Ci\`o significa che negli anni del suo ``isolamento", egli si
isol\`o dal gruppo di Fermi, s\`{\i}, ma non dalla vita: per lo meno, non
dalla vita scientifica e accademica, n\'e dalla fisica.

\h Non possiamo dimenticare gli appunti autografi di lezione redatti da
Majorana nei primi mesi del 1938 a beneficio dei propri studenti
dell'universit\`a di Napoli (probabilmente Ettore stava pensando di scrivere
un libro per studenti avanzati di fisica teorica). L'intera serie di appunti
per le sedici
lezioni da lui tenute prima della scomparsa fu consegnata dal Majorana,
entro una cartelletta, il giorno prima di scomparire, all'allieva
Gilda Senatore e (essendone intermediari Cennamo, Carrelli e Amaldi) fin\'{\i}
nelle mani di Gilberto Bernardini, probabilmente soltanto in parte, e quindi
negli archivi della ``Domus Galilaeana". La parte cos\'{\i} sopravvissuta
(relativa a dieci lezioni) fu pubblicata\cite{biblio} per interessamento di
di Gialanella
e soprattutto Bruno Preziosi, in un volume (Bibliopolis; Napoli, 1987)
contenente anche gli appunti per la prolusione al corso ---la lezione
inaugurale--- rinvenuti da Recami.  Recentissimamente S.Esposito,
in collaborazione con Antonico Drago e altri, ha scoperto gli appunti delle
restanti sei lezioni: e quindi l'intera serie \`e stata pubblicata quest'anno,
2006, presso il medesimo Editore napoletano.

\

\h Menzioniamo che nel 2003 S.Esposito, E.Majorana jr, A. ven der
Merwe e E.Recami hanno finalmente pubblicato per i tipi della
Kluwer Academic Press (Dordrecht e Boston), nel volume ``Ettore
Majorana -- Notes on Theoretical Physics",  una prima parte degli
appunti lasciati inediti dal Nostro: e, precisamente, i quaderni
(noti come i {\em Volumetti\/}), che comprendono i suoi appunti di
studio redatti in Roma tra il 1927, anno in cui abbandon\`o gli
studi di Ingeneria per passare a quelli di Fisica, e il 1931. Tale
volume, apparso in traduzione inglese, \`e uscito nella versione
originale italiana (a cura di S.Esposito e E.Recami) presso la
casa editrice Zanichelli di Bologna quest'anno, mel 2006. Si
potr\`a verificare come tali manoscritti siano un modello non solo
di ordine, divisi come erano e sono in argomenti e persino muniti
di indici, ma anche di originalit\`a, scelta dell'essenziale, e
sinteticit\`a; tanto che essi potrebbero venire riguardati, da un
lato, come un eccellente complemento ---dopo oltre settanta
anni--- di un testo {\em moderno} di fisica teorica, e,
dall'altro, come una miniera di nuovi spunti e idee teoriche, in
fisica e matematica, stimolanti e utili anche per la ricerca
scientifica contemporanea. Un futuro, secondo volume,
pubblicher\`a presto ---si spera---  almeno una frazione di altri
manoscritti inediti, ancora pi\`u tecnici, ma ancora pi\`u ricchi
di spunti scientifici originali: i cosiddetti {\em Quaderni},
contenenti le note scritte da Majorana durante le sue ricerche
scientifiche; mentre i fogli sparsi, pur interessanti (alcune
pagine originali sono state donate al Centro Ettore Majorana di
Erice (TP)), difficilmente vedranno la luce in tempi brevi.

\

\noindent{Esistono altri manoscritti di Majorana?}

Tornando alla lettera del 18 febbraio al padre, in essa abbiamo trovato la
notizia molto interessante che Ettore stava per pubblicare in tedesco,
estendendolo, l'ultimo suo articolo apparso sul ``Nuovo Cimento". Come sappiamo,
questo progetto non verr\`a poi realizzato; ma \`e importante ricordare
ancora una volta come Ettore
avesse in mente di generalizzare il lavoro in cui aveva introdotto la sua
equazione a infinite componenti. \ Anzi, la questione diviene del massimo rilievo
quando si leggano le lettere\cite{ER} inviate in quel periodo al Consiglio Nazionale
delle Ricerche (ritrovate presso gli archivi del C.N.R., e a noi pervenute
attraverso la cortesia di G.Fioravanti e soprattutto del collega M.De Maria).
Nella prima (21.1.33) Ettore specifica: $<<$Attendo attualmente
alla elaborazione di una teoria per la descrizione di particelle con momento
intrinseco arbitrario che ho iniziata in Italia, e di cui ho dato notizia
sommaria nel {\rm Nuovo Cimento} (in corso di stampa)...$>>$. \ Nella seconda
(3.3.33) dichiara addirittura, riferendosi al medesimo lavoro: $<<$Ho
inviato alla {\rm Zeitschrift f\"ur Physik} un articolo sulla teoria dei
nuclei. Ho pronto il manoscritto di una nuova teoria delle particelle
elementari e lo invier\`o alla stessa rivista fra qualche giorno...$>>$.
Se ricordiamo che l'articolo qui considerato
come ``notizia sommaria" di una nuova teoria era gi\`a di altissimo livello,
si comprende come sarebbe di enorme interesse scoprire una copia della teoria
completa: la quale nel marzo 1933 aveva gi\`a assunto la forma di un manoscritto
compiuto, forse gi\`a dattiloscritto in lingua tedesca. Ma Ettore, ripetiamo,
non ne fece pi\'u nulla. Non dimentichiamo poi la citata lettera a Quirino
del 16.1.1936, la quale ci ha rivelato che successivamente Ettore continu\`o
a lavorare in fisica
teorica, occupandosi a fondo ---per lo meno--- di elettrodinamica quantistica.
 \ Dove sono finiti gli appunti, gli scritti, gli articoli relativi a tutta
questa attivit\`a?

\h Come abbiamo gi\`a segnalato, il giorno prima di salpare da Napoli
(e successivamente sparire), Ettore Majorana consegn\`{o} alla propria
studentessa Sig.na Gilda Senatore [ora Prof.ssa Senatore] una cartelletta di
fogli scientifici: contenente, tra l'altro, gli appunti di lezione manoscritti
dal Majorana per i suoi allievi; affinch\'e lei la conservasse. Tutto ci\`o
lo si \`e saputo in seguito ad
una approfondita ricerca effettuata nel 1990 da Bruno Russo\cite{ER2},
e successivamente confermata a voce dalla stessa Prof.ssa Senatore a chi
scrive, nonch\'e a Bruno Preziosi.

\h La cartelletta conteneva (oltre alle ``lezioni") delle note incomplete,
degli scritti conclusi, e articoli. Si hanno ragioni per credere che tale
cartelletta contenesse anche alcuni almeno dei risultati
del lavoro svolto da Majorana, in isolamento, tra la fine del 1933 e il 1938.
Tali risultati sarebbero di straordinaria importanza, come sappiamo,
per la stessa fisica teorica contemporanea, pi\`u ancora che per la storia della
fisica. Ma avvenne che la Sig.na Senatore parl\`{o} confidenzialmente dei
manoscritti avuti in pegno da Majorana a Francesco Cennamo, assistente del
direttore Antonio Carrelli, quando questi divenne suo marito.  Il dottor
Cennamo, di propria iniziativa, li mostr\`o a Carrelli, che li sequestr\`o.
E, per quanto a noi ora consta, essi si persero.

\h Molte altre idee di Ettore, quando non restarono nella sua mente,
hanno lasciato traccia nella memoria dei colleghi. Una delle testimonianze
pi\'u interessanti che abbiamo raccolto \`e di GianCarlo Wick. Da Pisa il
16 Ott. 1978 scrive a Recami: \ $<<$...Il contatto scientifico
tra me ed Ettore di cui le accenn\`o Segr\'e avvenne a Roma
in occasione del Congresso Volta (assai prima del soggiorno di
Majorana a Lipsia). La conversazione ebbe luogo in un ristorante, in
presenza di Heitler, e dunque senza lavagna n\'e formule scritte; ma
nonostante l'assenza di dettagli quello che Majorana descrisse a
parole era una ``teoria relativistica di particelle cariche di spin
zero basata sull'idea di quantizzazione dei campi'' (seconda
quantizzazione). Quando assai pi\'u tardi vidi il lavoro di
Pauli [Premio Nobel 1945] e Weisskopf [{\it Helv. Phys. Acta} {\bf 7} (1934)
709], rimasi assolutamente convinto che quello che Majorana aveva
descritto fosse la stessa cosa$\ldots$ $>>$.

\

\

{\bf Ringraziamenti}

L'autore \`e molto grato a Silvano Sgrignoli, nonch\'e a tutti gli Organizzatori del
Covegno di
Catania: A.Agodi, P.Castorina, F.Catara, E.Migneco, S.Lo Nigro,
F.Porto, E.Rimini, e in particolare A.Rapisarda.  Ringrazia inltre,
per costante incoraggiamento o collaborazione, Franco Bassani,
Salvatore Esposito, Ignazio Licata, Fabio Majorana e Bruno Preziosi

\

\

\h Aggiungiamo una breve {\em Bibliografia}. Lungi dall'essere
completa, essa si riferisce soltanto agli argomenti toccati in
questo articolo, tralasciando le pubblicazioni gi\`a
esplicitamente citate nel testo. \ [Cogliamo l'occasione per
ricordare, infine, che molto del materiale che precede \`e
protetto, e non pu\`o essere ulteriormente riprodotto senza il
permesso scritto di Fabio Majorana et al., in solido con l'autore
e con il publisher, Di Renzo, Roma, della IV edizione, 2002, del
libro di E.Recami].

\